\documentclass[11pt,fleqn]{article}

\usepackage{graphicx}

\usepackage{latexsym}

\usepackage{amsmath}
\usepackage{amsthm}
\usepackage{amssymb}
\usepackage{amsfonts}
\usepackage{color}

\newcommand{\dis}{\displaystyle}
\newcommand{\rf}[1]{(\ref{#1})}

\newcommand{\ba}{\begin{array}}
\newcommand{\ea}{\end{array}}

\newcommand{\be}{\begin{equation}}
\newcommand{\ee}{\end{equation}}

\newcommand{\no}{\noindent}

\newcommand{\1}{{\pmb 1}}
\newcommand{\I}{{\pmb i}}

\newcommand{\ods}{\par \vspace{0.5cm} \par}

\newcommand{\oo}{\overset{o}}

\newenvironment{Proof}{\par \vspace{2ex} \par
\noindent  {\it Proof:}}{\hfill $\qedsymbol$ 
\vspace{2ex} \par }

\newtheorem{prop}{Proposition}[section]
\newtheorem{Th}[prop]{Theorem}
\newtheorem{lem}[prop]{Lemma}
\newtheorem{rem}[prop]{Remark}
\newtheorem{cor}[prop]{Corollary}
\newtheorem{Def}[prop]{Definition}

\numberwithin{equation}{section}

\begin{document}

\title{\bf  On geometry of the scator space}
\author{\bf  Artur Kobus
\\ {\footnotesize  Politechnika Bia\l ostocka, Wydzia{\l} Budownictwa i In\.zynierii \'Srodowiska}
\\ {\footnotesize  ul.\  Wiejska 45E,  15-351 Bia\l ystok, Poland}
\\[2ex] {\bf Jan L.\ Cie\'sli\'nski}\thanks{E-mail: \ 
 j.cieslinski@uwb.edu.pl}, 
\\ {\footnotesize Uniwersytet w Bia\l ymstoku,
Wydzia{\l} Fizyki}
\\ {\footnotesize ul.\ Cio\l kowskiego 1L,  15-245 Bia\l ystok, Poland}
}

\date{ }

\maketitle

\begin{abstract}
We consider the scator space - a hypercomplex, non-distributive hyperbolic algebra introduced by Fern\'andez-Guasti and Zald\'ivar. We discuss isometries of the scator space and find consequent method for treating them algebraically, along with scators themselves. It occurs that introduction of zero divisors cannot be avoided while dealing with these isometries. The scator algebra may be endowed with a nice physical interpretation, although it suffers from lack of some physically demanded important features. Despite that, there arises some open questions, e.g., whether hypothetical tachyons can be considered as usual particles  possessing time-like trajectories.
\end{abstract}

\noindent {\it MSC 2010}:  30G35;  20M14
\par 
\noindent {\it Keywords}: \par \no scators, tachyons, non-distributive algebras, hypercomplex numbers.
\par 

\section{Introduction}

Following  Fern\'andez-Guasti and Zald\'ivar  \cite{FZ1} we consider a commutative, non-distributive  $1+2$ dimensional algebra $S$, which is also associative provided that divisors of zero are excluded. The elements of this algebra will be called 
 $1+2$ dimensional scators \cite{FZ1}. Scators  (a kind of hypercomplex numbers) are denoted by $\oo{a} = (a_0;a_1,a_2)$, where components $a_1, a_2$   are referred to as \emph{director} components, and $a_0$ is usually called \emph{scalar} component, or, in physical context, \emph{temporal} component.   
The space of scators possesses the additive structure of usual vector space, with scalars being its subset, closed under addition and multiplication.  In this paper we confine ourselves to this definition of a scator,  leaving aside earlier concepts of scators as  objects generalizing scalars and vectors, see \cite{Ch,T}. 

It was shown in \cite{F1} that this algebra may be given physical interpretation corresponding to some ideas of special relativity, although metric in the scator space is different  from the standard metric of  Minkowski space. This scator metric (defined below) is called scator-deformed Lorentz metric.
It was emphasized that  scators and deformed metrics  can  describe a kinematics of some kind of particles, including  hypothetical tachyons. 

In our paper we study  metric properties of scators,  paying  special attention to proper definitions for causal realms appearing in this framework, what finally will lead to some convergence with work of Kapu\'scik \cite{K}.

We  emphasize  the fact that multiplication acts in a non-distributive way, what is the hallmark of the structure. Indeed, we  have
\begin{eqnarray}
\lefteqn{\overset{o}{a} \overset{o}{b} = (a_0; a_1, a_2) (b_0; b_1, b_2) ={}}\nonumber\\
& &{} = \left( a_0 b_0 + a_1 b_1 + a_2 b_2 + \frac{a_1 a_2 b_1 b_2}{a_0 b_0};  a_0 b_1 + a_1 b_0 + \frac{a_1 a_2 b_2}{a_0} + \frac{a_2 b_1 b_2}{b_0},  \right.   {}\nonumber\\ 
& &{}    \left. a_0 b_2 + a_2 b_0 + \frac{a_1 a_2 b_1}{a_0} + \frac{a_1 b_1 b_2}{b_0}  \right) . \label{scator-product}
\end{eqnarray}
Therefore, computing 
\be
\label{del}
\Delta (\overset{o}{a}, \overset{o}{b}; \overset{o}{c}) := (\overset{o}{a} + \overset{o}{b}) \overset{o}{c} - \overset{o}{a} \overset{o}{c} - \overset{o}{b} \overset{o}{c} = \frac{(b_0 a_1 - a_0 b_1)(a_0 b_2 - b_0 a_2)}{a_0 b_0 (a_0 + b_0)} \left(\frac{c_1 c_2}{c_0}; c_2, c_1 \right),
\ee
where $\overset{o}{c} = (c_0; c_1, c_2)$, we easily see that in the generic case the righ-hand side does not vanish, i.e.,  
$(\oo{a} + \oo{b}) \oo{c}  \neq  \oo{a} \oo{c} + \oo{b} \oo{c}$  provided that  $b_0 a_1 \neq a_0 b_1$ and $a_0 b_2 \neq b_0 a_2$.  Hence, in general, the scator product is not distributive.
By the way, the scator appearing on the right-hand side of \rf{del} will be referred to as  dual to $\oo{c}$ (see Definition~\ref{def-dual} below). 

Many properties of the scator product \rf{scator-product}  were widely investigated in many contexts \cite{FZ1,FZ2,FZ3}, also physical \cite{F1}.
To gain some insight in possible physical interpretation of the scator algebra, we recall here some basic terminology from \cite{FZ1} and \cite{FZ3}, although modified a little:

\begin{Def}
The  modulus squared of a scator is given by
\be  \label{modulus}
\|\overset{o}{a}\|^2 = \overset{o}{a} \overset{o}{a}^{*} = a_0^2 \left(1 - \frac{a_1^2}{a_0^2} \right) \left(1 - \frac{a_2^2}{a_0^2} \right) = a_0^2 - a_1^2 - a_2^2 + \frac{a_1^2 a_2^2}{a_0^2}  .
\ee
\end{Def}

\begin{Def}
We say that a scator is time-like, if
\be
a_0^2 > a_1^2,\quad \textrm{and} \quad a_0^2 > a_2^2, \quad\textrm{or}\quad a_0^2 < a_2^2 ,\quad\textrm{and}\quad a_0^2 < a_1^2 ,
\ee
it is said to be space-like, if
\be
a_0^2 < a_1^2,\quad \textrm{and}\quad a_0^2 > a_2^2 \quad \textrm{or} \quad a_0^2 < a_2^2 ,\quad \textrm{and} \quad a_0^2 > a_1^2 ,
\ee
and it is light-like, when
\be
a_0^2 = a_1^2, \quad\textrm{or}\quad a_0^2 = a_2^2.
\ee
\end{Def}

The division proposed above is analogous to what is well known from special relativity. Note that in both cases  componentwise addition of scators or vectors does not preserve this division. 
On the other hand, the norm of a product of two scators is just a product of their norms \cite{FZ3}, which is very useful in this context. In this paper we present  new results concerning  the metric structure of the scator space,  extending results of  \cite{F1,FZ3}.

\begin{figure}
\begin{center}
\includegraphics[width=12cm]{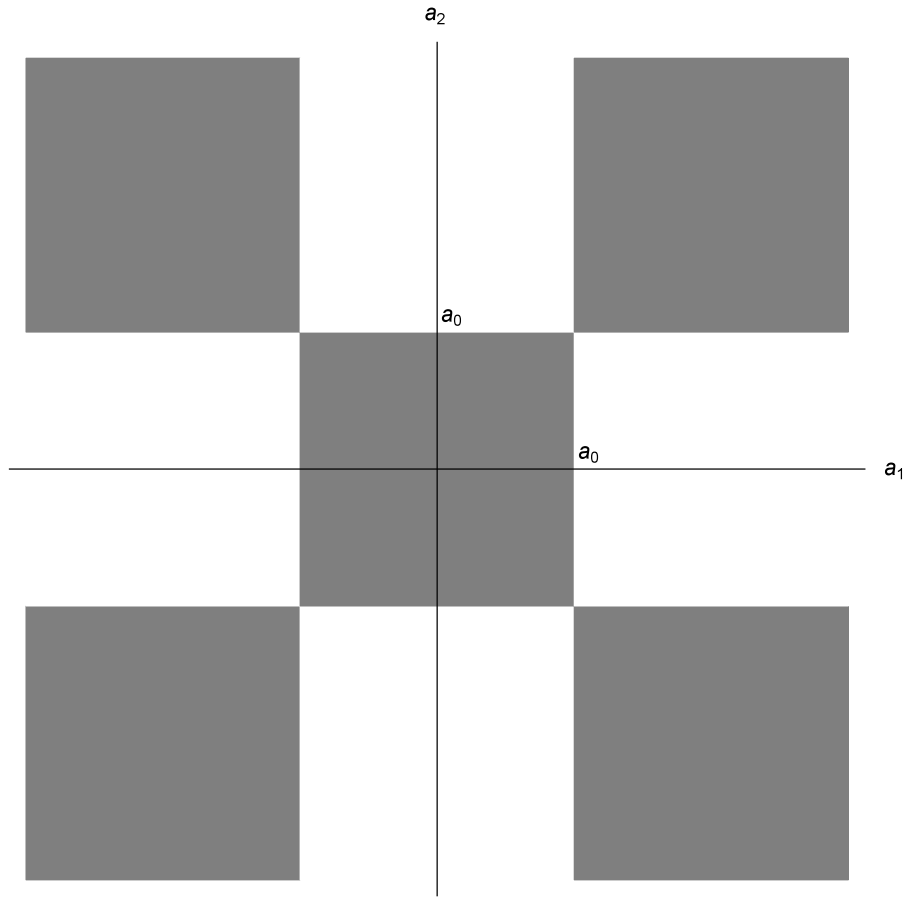}
\end{center}
\caption{Time-like wings at a fixed  time $a_0$ ($a_0 \neq 0$). Duality operations (section~\ref{sec-dual}) take us out from the bipyramid (represented in this cross-section by the inner square) to other causal realms, and conversely.}
\label{wings}
\end{figure}

For instance, we indicate that  bipyramid considered in  \cite{F1} has some kind of time-like ``wings''  around, described by the regime $a_0^2 < a_1^2$ and $a_0^2 < a_2^2$, while, for example, in \cite{FZ3}, there were considered mainly time-like events inside the light bipyramid ($a_0^2 > a_1^2$ and $a_0^2 > a_2^2$).  Time-like region is marked as dark at Fig.~\ref{wings}.

We underline that the existence of such wings has been overlooked in the paper \cite{F1}, what has its consequences in possible physical interpretation proposed in next sections. This seems to reveal some new aspects of causality in scator-deformed Lorentz metric (see the picture at end of the paper).

A closely related notion, ``super super-restricted space conditions'',  was introduced in \cite{FZ3} without a direct relation to the type of considered events. Super-restricted space conditions define either time-like events (in even-dimensional spaces) or space-like events (in odd-dimensional spaces).

The paper is organized as follows. In section $2$ we introduce  basic objects and transformations responsible for isometries in the scator space $S$. Then, in sections $3$ and $4$  we  propose and develop a new framework in which  calculation proceed  in a more natural way, using a distributive product.  In section  $5$ we continue  along these lines focusing on isometries.  Section $6$ is entirely devoted to the question of metric properties of scators; in particular, we obtain possible closest analogue of scalar product we can get, although it is even not bilinear. The last section contains physical comments and conclusions.

\section{Dualities - phenomenological treatment}
\label{sec-dual}

Now we turn our attention to the issue of isometries in the scator space $S$. We begin with defining some operations  and then check their properties.

\begin{Def}  \label{def-dual}
If we have a 3-scator $\overset{o}{a} = (a_0; a_1, a_2)$, then we call the scator of the form
\be  \label{bar}
\overset{o}{\bar{a}} = \left(\frac{a_1 a_2}{a_0}, a_2, a_1   \right)
\ee
its dual (or ordinary dual) scator, and star denotes hypercomplex conjugate:  
\be  \label{star}
\oo{a^*} = (a_0; - a_1, -a_2) .
\ee
\end{Def}

\begin{lem}
Operation of duality commutes  with hypercomplex conjugation.
\end{lem}

\no\emph{Proof:} We have
\be
(\overset{o}{\bar{a}})^* = \left(\frac{a_1 a_2}{a_0}, a_2, a_1\right)^* = \left(\frac{a_1 a_2}{a_0}, -a_2, -a_1\right)
\ee
and
\be
\overset{o}{\overline{(a^*)}} = \overline{(a_0; -a_1, -a_2)} = \left(\frac{a_1 a_2}{a_0}, -a_2, -a_1\right)
\ee
which are evidently equal. \hfill\qedsymbol

\begin{lem}
Operation of duality is idempotent:  $\oo{\overline{(\bar a)}} = \oo{a}$.
\end{lem}

\no\emph{Proof:} This follows instantly by straightforward calculation.  \hfill\qedsymbol

\begin{rem} Hypercomplex conjugation is a homomorphism in $S$, i.e., 
\be
    (\oo{a} \oo{b})^*  = (\oo{a})^* (\oo{b})^* \ .   
\ee
Operation of duality does not provide yet another homomorphic sctructure in $S$,  i.e.  
$\overset{o}{\overline{a b}} \neq \overset{o}{\bar{a}} \overset{o}{\bar{b}}$. 
\end{rem}

The above statements follow from considerations  similar to direct calculations  included in \cite{CK}. Soon we will get a better understanding of these facts by applying a new approach which is both faster and simpler. 

\begin{prop}
Operation of duality  preserves the scator product:  \
\be
\oo{\bar{a}} \oo{\bar b} = \oo{a} \oo{b} .
\ee
\end{prop}

\no\emph{Proof:} We present explicit calculation for the scalar component:
\be
(\overset{o}{a} \overset{o}{b})_0 = a_0 b_0 + a_1 b_1 + a_2 b_2 + \frac{a_1 a_2 b_1 b_2}{a_0 b_0},
\ee
and 
\begin{eqnarray}
\lefteqn{(\overset{o}{\bar{a}} \overset{o}{\bar{b}})_0 = \bar{a}_0 \bar{b}_0 + \bar{a}_1 \bar{b}_1 + \bar{a}_2 \bar{b}_2 + \frac{\bar{a}_1 \bar{a}_2 \bar{b}_1 \bar{b}_2}{\bar{a}_0 \bar{b}_0} ={}}\nonumber\\ & &{} = \frac{a_1 a_2 b_1 b_2}{a_0 b_0} + a_2 b_2 + a_1 b_1 + \frac{a_1 a_2 b_1 b_2}{\frac{a_1 a_2 b_1 b_2}{a_0 b_0}} ={}\nonumber\\ & &{} = a_0 b_0 + a_1 b_1 + a_2 b_2 + \frac{a_1 a_2 b_1 b_2}{a_0 b_0} = 
(\overset{o}{a} \overset{o}{b})_0 . 
\end{eqnarray}
Similar direct computation can be done for director components.   \hfill\qedsymbol

\begin{prop}
Operation of duality is an isometry in 3-scator space.
\end{prop}

\no\emph{Proof:} For ordinary scator $\overset{o}{a}$ we have its norm
\be
\|\overset{o}{a}\|^2 = \overset{o}{a} \overset{o}{a}^{*} = a_0^2 \left(1 - \frac{a_1^2}{a_0^2} \right) \left(1 - \frac{a_2^2}{a_0^2} \right).
\ee
Thus, for the dual scator $\overset{o}{\bar{a}}$, we get
\begin{eqnarray}
\lefteqn{ \|\overset{o}{\bar{a}}\|^2 = \overset{o}{\bar{a}} \overset{o}{\bar{a}}^* = \left(\frac{a_1 a_2}{a_0}, a_2, a_1\right)\left(\frac{a_1 a_2}{a_0}, -a_2, -a_1\right) = {}}\nonumber\\
& &{} = \frac{a_1^2 a_2^2}{a_0^2} \bigg(1 - \frac{a_2^2}{\frac{a_1^2 a_2^2}{a_0^2}} \bigg) \bigg(1 - \frac{a_1^2}{\frac{a_1^2 a_2^2}{a_0^2}} \bigg) ={}\nonumber\\
& &{} = \frac{a_1^2 a_2^2}{a_0^2} + a_0^2 - a_1^2 - a_2^2 =  a_0^2 \left(1 - \frac{a_1^2}{a_0^2} \right) \left(1 - \frac{a_2^2}{a_0^2} \right), \;
\end{eqnarray}
which exactly coincides with the norm of  the original scator.\hfill\qedsymbol

\begin{rem}  \label{rem-real}
Taking into account \rf{scator-product}, \rf{star} and \rf{bar},  we can easily verify that 
\be 
  \oo{a^*} \oo{b} + \oo{a} \oo{b^*} = 2 \left(   a_0 b_0 - a_1 b_1 - a_2 b_2 + \frac{a_1 a_2 b_1 b_2}{a_0 b_0} \right)    ,  
 \ee
where the right-hand side is  proportional to $\1$ (omitted for simplicity here and in many other places). 
\end{rem}

\begin{Def}
If we have a 3-scator $\overset{o}{a} = (a_0; a_1, a_2)$, then we call a scator of the form
\be
\overset{o}{\bar{a}}_i = \left(a_1; a_0, \frac{a_1 a_2}{a_0} \right)
\ee
its internal dual scator, and a scator of the form
\be
\overset{o}{\bar{a}}_e = \left(a_2; \frac{a_1 a_2}{a_0}, a_0 \right)
\ee
its external dual.
\end{Def}

\begin{lem}
Internal and external duality operations  anti-commute with hypercomplex conjugation.
\end{lem}

\no\emph{Proof:}  We have
\be
\overset{o}{(\bar{a}_i)}^* = \left(a_1; a_0, \frac{a_1 a_2}{a_0}\right)^* = \left(a_1; -a_0, -\frac{a_1 a_2}{a_0}\right)
\ee
and
\be
\overset{o}{\overline{(a^*)}}_i = \overline{(a_0; -a_1, -a_2)_i} = \left(-a_1; a_0, \frac{a_1 a_2}{a_0} \right) ,
\ee
so that
\be
(\overset{o}{\bar{a}})_i^* + \overset{o}{\overline{(a^*)}_i} = 0.
\ee
Similarly we have 
\be
(\overset{o}{\bar{a}})_e^* = (a_2; \frac{a_1 a_2}{a_0}, a_0)^* = \left(a_2; -\frac{a_1 a_2}{a_0}, -a_0\right)
\ee
and
\be
\overset{o}{\overline{(a^*)}}_e = \overline{(a_0; -a_1, -a_2)_e} =\left(-a_2; \frac{a_1 a_2}{a_0}, a_0\right) ,
\ee
so that
\be
(\overset{o}{\bar{a}}_e)^* + \overline{(a^*)_e} = 0,
\ee
which ends the proof. \hfill\qedsymbol

\begin{lem}
Internal and external duality operations are idempotent.
\end{lem}

\no\emph{Proof:} It is enough to apply twice definitions of both operations. \hfill\qedsymbol

\begin{rem}
Operation of external and internal duality do not provide yet another homomorphic sctructures in $S$, so i.e.
\be
(\overset{o}{\overline{a b}})_i \neq \overset{o}{\bar{a}}_i \overset{o}{\bar{b}}_i, \qquad (\overset{o}{\overline{a b}})_e \neq \overset{o}{\bar{a}}_e \overset{o}{\bar{b}}_e ,
\ee
which follows from straightforwad calculation. 
\end{rem}

\begin{lem}
Operations of internal and external duality preserve scator product.
\end{lem}

\no\emph{Proof:} By direct computation, similar to the case of ordinary duality. \hfill\qedsymbol

\begin{Def}
A transformation that exchanges time-like events with space-like events (and conversely) and leaves the type of  light-like events  unchanged  is called a causality swap (or a pseudo-isometry).
\end{Def}

\begin{prop}
Both internal and external duality operations are causality swaps of 3-scator space.
\end{prop}

 \no\emph{Proof:} Denoting $\oo{\bar a}_i = ({\bar a}_{0i}; {\bar a}_{1i}, {\bar a}_{2i})$, we compute:
\begin{eqnarray}
\lefteqn{\|\overset{o}{\bar{a}}_i\|^2 = \bar{a}_{0i}^2  \left(1 - \frac{\bar{a}_{1i}^2}{\bar{a}_{0i}^2}\right) \left(1 - \frac{\bar{a}_{2i}^2}{\bar{a}_{0i}^2} \right) = {}}\nonumber\\
& &{}= \bar{a}_{0i}^2 \left(1 - \frac{\bar{a}_{1i}^2}{\bar{a}_{0i}^2} - \frac{\bar{a}_{2i}^2}{\bar{a}_{0i}^2} + \frac{\bar{a}_{1i}^2}{\bar{a}_{0i}^2}\frac{\bar{a}_{2i}^2}{\bar{a}_{0i}^2}\right) = a_1^2 \left(1 - \frac{a_0^2}{a_1^2}\right) \left(1 - \frac{a_2^2}{a_1^2}\right) ={}\nonumber\\
& &{} = - a_0^2 \left(1 - \frac{a_1^2}{a_0^2}\right) \left(1 - \frac{a_2^2}{a_0^2} \right) = - \|\overset{o}{a}\|^2 \;
\end{eqnarray}
Therefore,  if original scator represents a time-like event, then its internal dual has to represent a space-like event, and {\it vice versa}. Light-like scators do not change their type. Analogous computation, with the same consequences,  can be done for the  external dual.  \hfill\qedsymbol

\begin{cor}
Duality operations commuting with hypercomplex conjugation are isometries, while duality operations anti-commuting with hypercomplex conjugation are causality swaps.
\end{cor}

Finally, we arrive at a very strong theorem providing some kind of translator between different kinds of duals.

\begin{Th}
\label{dual}
Ordinary, internal, and external duality operations have the following properties:
\begin{eqnarray}
\lefteqn{ \overset{o}{\bar{a}} \overset{o}{\bar{b}}_i = \overset{o}{\bar{a}}_i \overset{o}{\bar{b}} = (\overset{o}{\overline{a b}})_e = \overset{o}{(\bar{a}})_e \overset{o}{b} = \overset{o}{a} (\overset{o}{\bar{b}})_e,{} }\nonumber\\ & & {} \overset{o}{\bar{a}} \overset{o}{\bar{b}}_e = \overset{o}{\bar{a}}_e \overset{o}{\bar{b}} = (\overset{o}{\overline{a b}})_i =\overset{o}{(\bar{a}})_i \overset{o}{b} = \overset{o}{a} (\overset{o}{\bar{b}})_i {}\nonumber\\& & {}\overset{o}{\bar{a}}_e \overset{o}{\bar{b}}_i = \overset{o}{\bar{a}}_i \overset{o}{\bar{b}}_e = \overset{o}{\overline{a b}} =\overset{o}{\bar{a}} \overset{o}{b} = \overset{o}{a} \overset{o}{\bar{b}}.\;
\end{eqnarray}
\end{Th}

\no\emph{Proof:} 
One can perform lenghty straightforward calculation. However, applying a new approach of next sections we will be able to present a very short  proof. 
\hfill\qedsymbol

\section{ Advances in algebra}

Let $S'$ denotes the space of  hyperbolic 3-scators with non-vanishing scalar component ($\oo{a} \in S'$ if and only if $a_0 \neq 0$). 
We are going to  identify  $S'$  with  objects  in a linear space  $A$  of dimension 4, where the fourth component is reserved for some geometric invariant.  

\begin{Def}  The fundamental embedding of $S'$ into $A$ is given by the map 
$F: S'  \rightarrow \tilde S \equiv F(S) \subset A$,  defined by
\be  \label{fund}
S' \ni \oo{a} = (a_0; a_1, a_2) \rightarrow F (\oo{a}) :=(a_0; a_1, a_2, a_3) , \quad a_3 = \frac{a_1 a_2}{a_0} .
\ee
\end{Def}

\begin{rem}
$F$ is a bijection between $S'$ and $\tilde S$, so it has an inversion $F^{-1}: \tilde S \rightarrow S$.  We have also a natural projection
\be
  \pi : A \ni   (a_0; a_1, a_2, a_3) \rightarrow (a_0; a_1, a_2) \in S \ .
\ee
Note that  $\pi \neq F^{-1}$  but \ $\pi|_{\tilde S} = F^{-1}$. 
\end{rem}

We denote by $\{ \1, \pmb{i}_1, \pmb{i}_2, \pmb{i}_{12}\}$  a basis in the linear space $A$ such that 
\be
    (a_0; a_1, a_2, a_3) = a_0 \1 + a_1 \I_1 + a_2 \I_2  + a_3 \I_{12} .
\ee
The first three elements span the scator space $S$.  
The basis   $\{\1, \I_1, \I_2 \}$  is related to  $\{ 1, {\hat e}_1, {\hat e}_2 \}$  used in  papers \cite{FZ1,FZ2}.  As in these papers, we demand  that
\be
\label{v1}
\1^2 = \pmb{i}_1^2 = \pmb{i}_2^2 = \pmb{i}_{12}^2 = 1,
\ee
but we define
\be  \label{v2}
    \I_{12} = \I_1 \I_2 = \I_2 \I_1 \ .
\ee
while the Authors of \cite{FZ1,FZ2}  make  \emph{different} assumptions: $\hat{e}_1 \hat{e}_2 = \hat{e}_2 \hat{e}_1 = 0$. The last equalities can be neatly interpreted in our framework because
\be
F^{-1} (\pmb{i}_1 \pmb{i}_2) = F^{-1} (\pmb{i}_{12}) = 0 \ . 
\ee
Now we make our fundamental  assumption about the space $A$. We assume that this is a commutative, associative and \emph{distributive} algebra,  compare  \cite{CK}. We denote 
\be
F(\overset{o}{a}) = \left(a_0; a_1, a_2, \frac{a_1 a_2}{a_0}\right),\qquad F(\overset{o}{b}) = \left(b_0; b_1, b_2, \frac{b_1 b_2}{b_0}\right),
\ee
where we have written fourth components of scators explicitly.  in order to keep track of possible agreement with expected results.
We have
\[
F(\overset{o}{a}) F(\overset{o}{b}) = \left(a_0 + a_1 \pmb{i}_1 + a_2 \pmb{i}_2 + \frac{a_1 a_2}{a_0} \pmb{i}_{12} \right)\left(b_0 + b_1 \pmb{i}_1 + b_2 \pmb{i}_2 + \frac{b_1 b_2}{b_0} \pmb{i}_{12}\right) .
\]
Next, we use distributivity and (\ref{v1})
\begin{eqnarray}
\lefteqn{F(\overset{o}{a}) F(\overset{o}{b}) = a_0 b_0 + a_1 b_1 + a_2 b_2 + \frac{a_1 a_2 b_1 b_2}{a_0 b_0} + \pmb{i}_1 (a_0 b_1 + a_1 b_0) + {}}\nonumber\\ & & {}+ \pmb{i}_{12} \pmb{i}_1 \frac{a_1 a_2}{a_0} b_1 + \pmb{i}_1 \pmb{i}_{12} \frac{b_1 b_2}{b_0} a_1 + \pmb{i}_2 (a_0 b_2 + a_2 b_0) + \pmb{i}_{12} \pmb{i}_2 \frac{a_1 a_2}{a_0} b_2 +{}\nonumber\\ & & {}+ \pmb{i}_2 \pmb{i}_{12} \frac{b_1 b_2}{b_0} a_2 + \pmb{i}_{12} \left(\frac{a_1 a_2}{a_0} b_0 + \frac{b_1 b_2}{b_0} a_0\right) + \pmb{i}_1 \pmb{i}_2 (a_1 b_2 + a_2 b_1) .\;
\end{eqnarray}
Then, due to (\ref{v2}), we obtain
\begin{eqnarray}
\lefteqn{F(\overset{o}{a}) F(\overset{o}{b}) = a_0 b_0 + a_1 b_1 + a_2 b_2 + \frac{a_1 a_2 b_1 b_2}{a_0 b_0} + \pmb{i}_1 (a_0 b_1 + a_1 b_0) +{} }\nonumber\\ & & {}+ \pmb{i}_{12} \pmb{i}_2 \frac{a_1 a_2}{a_0} b_2 + \pmb{i}_2 \pmb{i}_{12} \frac{b_1 b_2}{b_0} a_2 + \pmb{i}_2 (a_0 b_2 + a_2 b_0)+ \pmb{i}_{12} \pmb{i}_1 \frac{a_1 a_2}{a_0} b_1+{} \nonumber \\ & & {} + \pmb{i}_1 \pmb{i}_{12} \frac{b_1 b_2}{b_0} a_1 + \pmb{i}_{12} \left(\frac{a_1 a_2}{a_0} b_0 + \frac{b_1 b_2}{b_0} a_0 + a_1 b_2 + a_2 b_1\right).\;
\end{eqnarray}
We see that the obtained scalar component coincides with  the scalar component of \rf{scator-product}. In order to get the same director components we have to assume 
\be
\pmb{i}_1 \pmb{i}_{12} = \pmb{i}_{12}\pmb{i}_1 = \pmb{i}_2, \qquad \pmb{i}_2 \pmb{i}_{12} = \pmb{i}_{12}\pmb{i}_2 = \pmb{i}_1, 
\ee
which follows from \rf{v1} and \rf{v2}. Finally, we get
\begin{eqnarray}
\label{res}
\lefteqn{F(\overset{o}{a}) F(\overset{o}{b}) = a_0 b_0 + a_1 b_1 + a_2 b_2 + \frac{a_1 a_2 b_1 b_2}{a_0 b_0} + \pmb{i}_1 (a_0 b_1 + a_1 b_0 +   {} } \nonumber\\ & & {}  + \frac{a_1 a_2}{a_0} b_2   + \frac{b_1 b_2}{b_0} a_2  )  + \pmb{i}_2 (a_0 b_2 + a_2 b_0 + \frac{a_1 a_2}{a_0} b_1 +{} \nonumber \\ & & {} + \frac{b_1 b_2}{b_0} a_1) + \pmb{i}_{12} \left(\frac{a_1 a_2}{a_0} b_0 + \frac{b_1 b_2}{b_0} a_0 + a_1 b_2 + a_2 b_1 \right) .\;
\end{eqnarray}

\begin{cor}
From \rf{res} we immediately see that  \ $\oo{a} \oo{b} = \pi ( F (\oo{a}) F (\oo{b}) )$. 
\end{cor}
\ods

\begin{Th}  \label{Th-fund}
The fundamental embedding \rf{fund}   is a multiplicative homomorphism of $S$ and $\tilde S$: 
\be  \label{fund-prod}
F(\overset{o}{a}) F(\overset{o}{b}) = F(\overset{o}{a} \overset{o}{b}), \quad   \text{hence}  \quad   \overset{o}{a} \overset{o}{b} = F^{-1} (F(\overset{o}{a}) F(\overset{o}{b})). 
\ee
\end{Th}

\begin{Proof}
We may check it by tedious straightforward computation, multiplying scalar and fourth component of \rf{res} and comparing it with the product of director components. 
However, calculations can be avoided  when we take into account the following factorization:
\[   \ba{l} \dis
F(\oo{a}) = a_0 \left( 1 + \frac{a_1}{a_0} \I_1 \right) \left( 1 + \frac{a_2}{a_0} \I_2  \right) ,  \quad 
\dis   F(\oo{b}) = b_0 \left( 1 + \frac{b_1}{b_0} \I_1 \right) \left( 1 + \frac{b_2}{b_0} \I_2  \right)   ,
\ea \]
\[
 F(\oo{a}) F (\oo{b}) = a_0 b_0  \left(  1 + \frac{a_1 b_1}{a_0 b_0} +  \left(  \frac{a_1}{a_0} + \frac{b_1}{b_0} \right)  \I_1 \right)  \left(  1 + \frac{a_2 b_2}{a_0 b_0} + \left(  \frac{a_2}{a_0} + \frac{b_2}{b_0}  \right)   \I_2 \right) .
\]
We see immediately that $ F(\oo{a}) F (\oo{b}) \in \tilde S$ which completes the proof. 
\end{Proof}

\begin{rem}  We can easily check that
\be
F(\overset{o}{a^*}) = ( F(\overset{o}{a}) )^* , \quad              F(\lambda \overset{o}{a}) = \lambda F(\overset{o}{a}), \quad 
F^{-1} (\lambda F (\oo{a})) = \lambda \oo{a} , 
\ee
where $\lambda$ is a real  constant.
\end{rem}

\section{Formula for distributive multiplication}

A crucial point in our analysis is to express the difference \rf{del}  in terms of the fundamental embedding.  We take into account  distributivity of the algebra $A$ assumed in the previous section.  Note that although $F$ is a multiplicative homomorphism but is not additive, i.e.,  in general
\be
F(\overset{o}{a}) + F(\overset{o}{b}) \neq F(\overset{o}{a} + \overset{o}{b}).
\ee
Therefore,  multiplication of scators has a  geometric interpretation, while addition of such objects cannot be treated geometricly.
In particular, 
\be
(F (\overset{o}{a} + \overset{o}{b})) F (\overset{o}{c}) \neq F (\overset{o}{a}) F (\overset{o}{c}) + F(\overset{o}{b}) F (\overset{o}{c}) .
\ee
But then we surely have
\be
F (\overset{o}{a}) F (\overset{o}{c}) + F(\overset{o}{b}) F (\overset{o}{c}) = (F (\overset{o}{a}) + F(\overset{o}{b})) F (\overset{o}{c}),
\ee
because the product in $\tilde S$ \emph{is} distributive.  
We will try express $F (\oo{a} + \oo{b})$ in terms of  $F (\oo{a})$ and $F(\oo{b})$.    First, we  compute
\be \ba{l} \dis
F(\overset{o}{a} + \overset{o}{b}) F(\overset{o}{c}) = \\[3ex]\dis
{} = \left(a_0; a_1, a_2, \frac{a_1 a_2}{a_0}\right)\left(c_0; c_1, c_2, \frac{c_1 c_2}{c_0}\right) + \left(b_0; b_1, b_2, \frac{b_1 b_2}{b_0}\right) \left(c_0; c_1, c_2, \frac{c_1 c_2}{c_0}\right)    \\[3ex]\dis
 {}= \left(c_0 (a_0 + b_0) + c_1 (a_1 + b_1) + c_2 (a_2 + b_2) + \frac{(a_1 + b_1)(a_2 + b_2)}{a_0 + b_0} \frac{c_1 c_2}{c_0};  \right.  {}\\[3ex]\dis
 \qquad c_1 (a_0 + b_0) + c_0 (a_1 + b_1) + \frac{c_1 c_2}{c_0} (a_2 + b_2) + \frac{(a_1 + b_1)(a_2 + b_2)}{a_0 + b_0} c_2, {}  \\[3ex]\dis
\qquad c_2 (a_0 + b_0) + c_0 (a_2 + b_2) + \frac{c_1 c_2}{c_0} (a_1 + b_1) + \frac{(a_1 + b_1)(a_2 + b_2)}{a_0 + b_0} c_1, {}\\[3ex]\dis
\qquad  \left. c_1 (a_2 + b_2) + c_2 (a_1 + b_1) + \frac{c_1 c_2}{c_0} (a_0 + b_0) + \frac{(a_1 + b_1)(a_2 + b_2)}{a_0 + b_0} c_0  \right) .\;
\ea \ee
On the other hand
\be \ba{l} \dis
F(\overset{o}{a}) F(\overset{o}{c}) + F(\overset{o}{b}) F(\overset{o}{c}) =   \\[3ex]\dis
= \left(a_0; a_1, a_2, \frac{a_1 a_2}{a_0}\right)\left(c_0; c_1, c_2, \frac{c_1 c_2}{c_0}\right) + \left(b_0; b_1, b_2, \frac{b_1 b_2}{b_0}\right)\left(c_0; c_1, c_2, \frac{c_1 c_2}{c_0}\right)     \\[3ex]\dis
= \left(c_0 (a_0 + b_0) + c_1 (a_1 + b_1) + c_2 (a_2 + b_2) + \left(\frac{a_1 a_2}{a_0} \frac{b_1 b_2}{b_0}\right) \frac{c_1 c_2}{c_0}; \right.  \\[3ex]\dis
\qquad  c_1 (a_0 + b_0) + c_0 (a_1 + b_1) + \frac{c_1 c_2}{c_0} \left(a_2 + b_2\right) + \left(\frac{a_1 a_2}{a_0} + \frac{b_1 b_2}{b_0}\right) c_2, \\[3ex]\dis
\qquad  c_2 (a_0 + b_0) + c_0 (a_2 + b_2) + \frac{c_1 c_2}{c_0} \left(a_1 + b_1\right) + \left(\frac{a_1 a_2}{a_0} + \frac{b_1 b_2}{b_0}\right) c_1, \\[3ex]\dis
\qquad  \left.  c_1 (a_2 + b_2) + c_2 (a_1 + b_1) + \frac{c_1 c_2}{c_0} \left(a_0 + b_0\right) + \left(\frac{a_1 a_2}{a_0} + \frac{b_1 b_2}{b_0}\right) c_0 \right).\;
\ea \ee
Therefore 
\begin{eqnarray}   \label{46}
\lefteqn{F(\overset{o}{a} + \overset{o}{b}) F(\overset{o}{c}) - F(\overset{o}{a}) F(\overset{o}{c}) - F(\overset{o}{b}) F(\overset{o}{c}) ={}}\nonumber\\[2ex]
& & {}= \left(\frac{(a_1 + b_1)(a_2 + b_2)}{a_0 + b_0} - \frac{a_1 a_2}{a_0} - \frac{b_1 b_2}{b_0}\right)\left(\frac{c_1 c_2}{c_0}; c_2, c_1, c_0\right) \;
\end{eqnarray}
and this  is  exactly  $F (\Delta (a, b; c))$, as defined in (\ref{del})!  Thus we have
\be
\label{fff}
F((\overset{o}{a} + \overset{o}{b}) \overset{o}{c} - \overset{o}{a} \overset{o}{c} - \overset{o}{b} \overset{o}{c}) = F(\overset{o}{a} + \overset{o}{b}) F(\overset{o}{c}) - F(\overset{o}{a}) F(\overset{o}{c}) - F(\overset{o}{b}) F(\overset{o}{c}),
\ee
which looks suspiciously homomorphic.  Equation \rf{46} can be rewritten as 
\be  \label{48}
F(\overset{o}{a} + \overset{o}{b}) F(\overset{o}{c}) - F(\overset{o}{a}) F(\overset{o}{c}) - F(\overset{o}{b}) F(\overset{o}{c}) = \kappa  (\oo{a},\oo{b})  F(\overset{o}{\bar{c}})
\ee
where $\kappa  (\oo{a},\oo{b})$ is  the scalar function  standing before dual of $\overset{o}{c}$ in formulas  (\ref{del}) and \rf{46}.

\begin{rem}
From \rf{modulus} it follows that the inverse of a scator (with respect to the scator product)  is given by 
\be
   ( \oo{c})^{-1} = \frac{(\oo{c})^* }{\| \oo{c} \|^{2}} \ ,
\ee
and light-like scators are not invertible. 
\end{rem}

\begin{Th}  \label{Th-sum}
\be
\label{forg}
F(\overset{o}{a} + \overset{o}{b}) - F(\overset{o}{a}) - F(\overset{o}{b}) = \kappa  (\oo{a},\oo{b})  \I_{12}  .
\ee
\end{Th}

\begin{Proof}
We multiply both sides of  \rf{48} by $F ((\overset{o}{c})^{-1})$ and, taking into account that  $F$ is a homomorphism, we obtain the left-hand side of \rf{forg}. Then we observe that 
\be
   F(\overset{o}{\bar{c}}) = \frac{c_1 c_2}{c_0} + c_2 \I_1 + c_1 \I_2 + c_0 \I_{12} = \I_{12} F (\oo{c}) . 
\ee
Therefore
\be   \label{I12}
    F(\overset{o}{\bar{c}}) F ((\overset{o}{c})^{-1}) = \frac{\I_{12}  F ( (\oo{c})^*)  F (\oo{c})  }{\|\oo{c}\|^2 } = \I_{12} 
\ee
which ends the proof.
\end{Proof}

\begin{rem}  As a direct consequence of  \rf{I12} we get the following strange result
\be
\label{mj}
\overset{o}{\bar{a}} (\overset{o}{a})^{-1} = 0 ,
\ee
which manifestly shows that the scator algebra has numerous zero divisors.
\end{rem}

\begin{cor}
Theorem~\ref{Th-sum} implies that for a given set of scators, $\overset{o}{a}_i = (a_{0i}; a_{1i}, a_{2i})$, where $i = 1,\ldots, n$, we have: 
\be
F (\overset{o}{a}_1 + \ldots + \overset{o}{a}_n) = F (\overset{o}{a}_1) + F (\oo{a}_2) + \ldots +  F (\overset{o}{a}_n) + \kappa_n (\overset{o}{a}_1,  \oo{a}_2, \ldots,\overset{o}{a}_n) \pmb{i}_{12} \ .
\ee
where $\kappa_n$  (a symmetric scalar function) can be expressed by $\kappa$:
\be
  \kappa_n (\overset{o}{a}_1,  \oo{a}_2, \ldots,\overset{o}{a}_n)   =  \sum_{j=2}^n   \kappa (\oo{a_1}+ \oo{a}_2 + \ldots + \oo{a}_{j-1}, \oo{a}_{j}) \ .
\ee
\end{cor}

\begin{cor}
From Theorem~\ref{Th-sum} it follows that  the  inverse of the fundamental embedding is an additive homomorphism, i.e., 
\be
 F^{-1} ( F (\oo{a}) + F (\oo{b}) ) = F^{1} ( F (\oo{a} ) + F^{-1} ( F (\oo{b}) ) \equiv \oo{a} + \oo{b} ,
\ee
because $F^{-1} (\I_{12}) = 0$. 
\end{cor}

We point out that  $F^{-1}$  is not a multiplicative homomorphism. Counter-examples will be given in the next section.

\section{Dualities - systematic treatment}

Here we also provide some back-up for what we have done earlier: at first, we see why we call duality operation a \emph{duality} operation. First, we note that multiplication by bivector $\I_{12}$ acts on any element of the algebra $A$ in a way similar to the Hodge operator producing its complementary element with respect to the ``maximal form'' $\I_{12}$. We have two pairs of complementary basis elements: the first one is $\1$ and $\I_{12}$ and the second one is $\I_1$ and $\I_2$.

\begin{Def}
Duality operations are naturally given by:
\be  \label{def-dualities}
\overset{o}{\bar{a}} = F^{-1} (\pmb{i}_{12} F(\overset{o}{a})) , \quad
 \overset{o}{\bar{a}}_i = F^{-1} (\pmb{i}_1 F(\overset{o}{a})) , \quad
 \overset{o}{\bar{a}}_e = F^{-1} (\pmb{i}_2 F(\overset{o}{a})), 
\ee
and, of course,   $\overset{o}{a} = F^{-1} (\1   F(\overset{o}{a}))$.
\end{Def}

Therefore, we have
\be   \label{cor-dualities} 
F(\overset{o}{\bar{a}}) = \pmb{i}_{12} F(\overset{o}{a}) , \quad
 F(\overset{o}{\bar{a}}_i) = \pmb{i}_1 F(\overset{o}{a}), \quad
 F(\overset{o}{\bar{a}}_e) = \pmb{i}_2 F(\overset{o}{a}) .
 \ee
Thus, taking into account $F (\oo{a})^{-1}) = (F(\oo{a}))^{-1}$ and \rf{fund-prod}, we get: 
\be
\overset{o}{\bar{a}} \overset{o}{a}^{-1} = 0,\quad \overset{o}{\bar{a}}_i \overset{o}{a}^{-1} = \pmb{i}_1,\quad \overset{o}{\bar{a}}_e \overset{o}{a}^{-1} = \pmb{i}_2 .
\ee

In order to unify proofs it is convenient to introduce the following   notation, generalizing \rf{def-dualities} and \rf{cor-dualities}:
\be  \label{duals}
   \delta_{\pmb{d}} (\oo{a}) = F^{-1} ( \pmb{d} F (\oo{a}) ) , \quad  F (\delta_{\pmb{d}} (\oo{a})) = \pmb{d} F (\oo{a}) , \quad  F (\oo{a}) = \pmb{d} F (\delta_{\pmb{d}} (\oo{a})) , 
\ee
where $\delta_{\pmb{d}}$ denotes one of duality operations and $\pmb{d}$ denotes the corresponding basis element ($\I_1$, $\I_2$ or $\I_{12}$).  Note that $\pmb{d}^2 = 1$.

\begin{lem}
Any duality operation commutes with inversion, i.e., 
\be   \label{dual-inv}
\delta_{\pmb{d}} ((\oo{a})^{-1}) = (\delta_{\pmb{d}}(\oo{a}))^{-1} . 
\ee
\end{lem}
\begin{Proof}  Using  $F (\oo{a}) = \pmb{d} F (\delta_{\pmb{d}} (\oo{a}))$ and  $1 = F(\overset{o}{a}) F((\overset{o}{a})^{-1})$  (because $F$ is a homomorphism) we obtain 
\[
1 = \pmb{d} F (\delta_{\pmb{d}} (\oo{a})) F ((\oo{a})^{-1}) = F (\delta_{\pmb{d}} (\oo{a})) \pmb{d}   F ((\oo{a})^{-1}) =  F (\delta_{\pmb{d}} (\oo{a})) F (\delta_{\pmb{d}} ((\oo{a})^{-1})) .
\]
On the othe hand, we have  $1 = F(\delta_{\pmb{d}}(\oo{a})) F ((\delta_{\pmb{d}}(\oo{a}))^{-1})$. Therefore, 
\[
    F (\delta_{\pmb{d}} ((\oo{a})^{-1})) =    F ((\delta_{\pmb{d}}(\oo{a}))^{-1}) 
\]
and, taking into account that $F$ is a bijection, we obtain \rf{dual-inv}. 
\end{Proof}

Now, we can give a simple the proof for Theorem~\ref{dual}. We rewrite this theorem in a more convenient way, using the unification of dualities formulated in this section.  

\begin{Th}  \label{Th-pq}
Duality operations have the following properties: 
\be  \ba{l}
\delta_{\pmb{p}}  (\oo{a} \oo{b}) = \delta_{\pmb{p}} (\oo{a}) \oo{b} = \oo{a} \delta_{\pmb{p}} (\oo{b}) , \\[1ex]
 \delta_{\pmb{p}}  (\oo{a}) \delta_{\pmb{q}} (\oo{b}) = \delta_{\pmb{q}}  (\oo{a}) \delta_{\pmb{p}}  (\oo{b})  = \delta_{\pmb{p q}} (\oo{a} \oo{b}) ,
\ea \ee
where $\pmb{p}, \pmb{q} \in \{ \I_1, \I_2, \I_{12} \}$. 
\end{Th} 

\begin{Proof}:  Using \rf{duals} and \rf{fund-prod} we obtain:
\be
  \delta_{\pmb{p}}  (\oo{a} \oo{b}) = F^{-1} (\pmb{p} F (\oo{a}\oo{b}) ) = F^{-1} ( \pmb{p} F (\oo{a}) F (\oo{b}) ) =  F^{-1} (  F (\oo{a}) \pmb{p} F (\oo{b}) ) .  
\ee
\be  \ba{l}
 F^{-1} ( \pmb{p} F (\oo{a}) F (\oo{b}) )  = F^{-1} ( F (\delta_{\pmb{p}} (\oo{a}) ) F(\oo{b})) = \delta_{\pmb{p}} (\oo{a}) \oo{b} , \\[1ex]
 F^{-1} (  F (\oo{a}) \pmb{p} F (\oo{b}) ) = F^{-1} ( F (\oo{a}) F (\delta_{\pmb{p}} (\oo{b}) ) ) = \oo{a} \delta_{\pmb{p}} (\oo{b}) , 
\ea \ee
which ends the proof. 
\end{Proof}

\begin{cor}
The preservation of the scator product by any duality is a special case of Theorem~\ref{Th-pq}  for $\pmb{q} = \pmb{p}$, namely:   
\be
     \delta_{\pmb{p}} (\oo{a}) \delta_{\pmb{p}} (\oo{b}) = \oo{a} \oo{b} \ ,
\ee
because  $\delta_{\pmb{1}}$ is the identity operation. 
\end{cor}

\section{Metric properties of scators}

We begin witha  simple fact.

\begin{lem}
Modulus squared of a scator satisfies:
\be
\|\overset{o}{a} \overset{o}{b}\|^2 = \|\overset{o}{a} \|^2 \|\overset{o}{b}\|^2 \ , \qquad  \| \lambda \oo{a} \| = |\lambda| \| \oo{a} \|  .
\ee
where $\lambda$ is a real constant. 
\end{lem}

\begin{Proof}  Using the definition \rf{modulus} we have
\be
\|\overset{o}{a} \overset{o}{b}\|^2 = (\overset{o}{a} \overset{o}{b}) (\overset{o}{a} \overset{o}{b})^* = (\overset{o}{a} \overset{o}{a^*}) (\overset{o}{b} \overset{o}{b^*})  =  \|\overset{o}{a} \|^2 \|\overset{o}{b}\|^2 ,
\ee
by virtue of  commutativity and associativity of the scator product.   The second equality follows directly from \rf{modulus}. 
 \end{Proof}

Therefore, the modulus squared of the scator product factorizes like the product of complex numbers.  Unfortunatelly, the modulus squared is not a quadratic form of the scator components, which means that there is not corresponding bilinear form. However, we will introduce an analogue of the scalar product postulating the formula obeyed by quadratic forms. 

\begin{Def}  \label{def-scal}
Scalar product of scators is defined  by
\be
\overset{o}{a} \cdot \overset{o}{b} = \frac{1}{2} \left(\| \overset{o}{a} + \overset{o}{b}\|^2 - \|\overset{o}{a} \|^2 - \|\overset{o}{b}\|^2\right).
\ee
\end{Def}

\begin{cor}
\be   \label{sq}
\overset{o}{a} \cdot \overset{o}{a} = \|\overset{o}{a}\|^2.
\ee
\end{cor}

\begin{Th}
The scalar product in $S$ is explicitly  given  by
\be
\label{sp}
\overset{o}{a} \cdot \overset{o}{b} = a_0 b_0 - a_1 b_1 - a_2 b_2 + \frac{(a_1 + b_1)^2 (a_2 + b_2)^2}{2 (a_0 + b_0)^2} -  \frac{a_1^2 a_2^2}{2 a_0^2} -  \frac{b_1^2 b_2^2}{2 b_0^2}.
\ee
\end{Th}

\begin{Proof}
We start from
\be
\|\overset{o}{a} + \overset{o}{b}\|^2 = (\overset{o}{a} + \overset{o}{b}) (\overset{o}{a} + \overset{o}{b})^* = (\overset{o}{a} + \overset{o}{b}) (\overset{o}{a^*} + \overset{o}{b^*}),
\ee
and, since we cannot safely proceed because of lack of distributivity in $S$, we move on to $\tilde S$, i.e., 
\be
\|\overset{o}{a} + \overset{o}{b}\|^2 = F^{-1} (F(\overset{o}{a} + \overset{o}{b}) F(\overset{o}{a^*} + \overset{o}{b^*})),
\ee
and from the formula (\ref{forg}) we get
\[
\|\overset{o}{a} + \overset{o}{b}\|^2 = F^{-1} ((F(\overset{o}{a}) + F(\overset{o}{b}) + \kappa (\overset{o}{a},\overset{o}{b}) \pmb{i}_{12}) (F(\overset{o}{a^*}) + F(\overset{o}{b^*}) + \kappa (\overset{o}{a^*},\overset{o}{b^*}) \pmb{i}_{12})) .
\]
Hence, using \rf{cor-dualities} and Theorem~\ref{Th-fund}, we obtain
\begin{eqnarray}  \label{scal-middle}
\lefteqn{\|\overset{o}{a} + \overset{o}{b}\|^2 =  \oo{a} \oo{a^*} +  \oo{b} \oo{b^*} +  \oo{a} \oo{b^*} + \oo{a^*} \oo{b}  +  }\nonumber\\
& & {}  + \kappa (\overset{o}{a},\overset{o}{b}) ( \oo{\bar a^*} + \oo{\bar b^*} ) + \kappa (\overset{o}{a^*},\overset{o}{b^*})  ( \overset{o}{\bar{a}} + \overset{o}{\bar{b}})  + \kappa (\overset{o}{a},\overset{o}{b})  \kappa (\overset{o}{a^*},\overset{o}{b^*})  .  
\end{eqnarray}
To proceed further we need Remark~\ref{rem-real} and the following properties  of the hypercomplex conjugation and $\kappa$:  
\be  
\oo{\bar a^*} + \oo{\bar{a}} = \frac{2 a_1 a_2}{a_0} \ , 
\ee
\be \label{kappy}
 \kappa (\overset{o}{a^*},\overset{o}{b^*}) = \kappa (\overset{o}{a},\overset{o}{b})  = \frac{(a_1 + b_1)(a_2+b_2)}{(a_0+b_0)} - \frac{a_1 a_2}{a_0} - \frac{b_1 b_2}{b_0} .
\ee
 Then from  \rf{scal-middle} and Definition~\ref{def-scal}  we have
\[ \ba{l} \dis
2 \oo{a} \cdot \oo{b} =  \kappa^2 +  2 \kappa \left(      \frac{a_1 a_2}{a_0}  + \frac{b_1 b_2}{b_0}  \right)   
+   2 \left( a_0 b_0 - a_1 b_1 - a_2 b_2 + \frac{a_1 a_2 b_1 b_2}{a_0 b_0}  \right)  \\[3ex]
\dis  \qquad = \left( \kappa +    \frac{a_1 a_2}{a_0}  + \frac{b_1 b_2}{b_0} \right)^2  -  \frac{a_1^2 a_2^2}{a_0^2} -  \frac{b_1^2 b_2^2}{b_0^2} + 2(a_0 b_0 - a_1 b_1 - a_2 b_2 ) ,
\ea \]
where $\kappa = \kappa  (\oo{a}, \oo{b})$. We complete the proof by substituting \rf{kappy}.   \end{Proof}

\begin{rem}
We  easily see that
\be
   (\lambda \overset{o}{a}) \cdot (\lambda \overset{o}{b}) = \lambda^2 (\overset{o}{a} \cdot \overset{o}{b}) .
\ee
but, in general, $\lambda (\overset{o}{a} \cdot \overset{o}{b}) \neq (\lambda \overset{o}{a}) \cdot \overset{o}{b} \neq \overset{o}{a} \cdot (\lambda \overset{o}{b}) \neq \lambda (\overset{o}{a} \cdot \overset{o}{b})$ and 
\be
(\overset{o}{a} + \overset{o}{b}) \cdot \overset{o}{c} \neq \overset{o}{a} \cdot \overset{o}{c} + \overset{o}{b} \cdot \overset{o}{c}.
\ee
We point out that this scalar product is ill-defined if a scalar component of any factor is zero. 
\end{rem}

\section{Underlying  physics  and conclusions}

We  point out  that the scator metric \rf{modulus} approaches the flat Minkowski metric of  special relativity space-time in the limit $a_0 \rightarrow \infty$. It was shown in \cite{F1} that the framework of scators excludes possibility of existence of an absolute rest frame, so it is consistent with the principle of relativity \cite{AC}. Therefore the subject is not only of purely mathematical interest but may have interesting physical points, as well. 

The first question is, can we think of tachyons as para-particles possessing space-like trajectories?
In the scator framework this would mean that at each instance of time tachyons have to be described by scators with negative modulus squared.  Thus understood tachyons, in order to  remain tachyons, need to experience sub-luminal Lorentz boosts at that each instance of time, since otherwise the sign of the modulus squared  would get reverted.
It is hard to imagine inertial, super-luminal observer that could be turned into another one with some kind of sub-luminal Lorentz-like transformation preserving scator metric.
Hence it seems to us that  a more natural way of understanding tachyons is that they are ordinary particles  in a different causal domain and they cannot reach us because of the infinite-energy requirement to pass the bipyramidal light-barrier, inside which we are capable of taking the measurements. 

To sum up: we may suppose that tachyons are being super-luminal in a sense of belonging to a different sub-luminal causal realm, although then we look exactly the same way to them.
Note that this hypothesis is surprisingly in accordance with recent remarks on the nature of tachyons \cite{K}.

Next question that appears in the physical context  is whether the approach proposed in this paper works  in the case of $1+3$ dimensions?
The answer is \emph{yes}. It may be easily shown that in physical space-time of scators there is
\be
\label{homm}
F(\overset{o}{a} + \overset{o}{b}) = F(\overset{o}{a}) + F(\overset{o}{b}) + c_{1,2} \pmb{i}_1 \pmb{i}_2 + c_{1,3} \pmb{i}_1 \pmb{i}_3 + c_{2,3} \pmb{i}_2 \hat{e_3} + c_{1,2,3} \pmb{i}_1 \pmb{i}_2 \pmb{i}_3,
\ee
as opposed to (\ref{forg}), opening all doors needed \cite{CK}.
In the above expression we have
\be
c_{i,j} (\overset{o}{a}, \overset{o}{b}) = \frac{(a_i + b_i)(a_j + b_j)}{a_0 + b_0} - \frac{a_i a_j}{a_0} - \frac{b_i b_j}{b_0}
\ee
and
\be
c_{1,2,3} (\overset{o}{a}, \overset{o}{b}) = \frac{(a_1 + b_1)(a_2 + b_2)(a_3 + b_3)}{(a_0 + b_0)^2} - \frac{a_1 a_2 a_3}{a_0^2} - \frac{b_1 b_2 b_3}{b_0^2} .
\ee
We see that the generalization, although simple in principle, may lead to cumbersome calculations. 
This also implies existence of more dualities, associated with  the basis of the  $A$-space, now $8$-dimesional.
Here also dualities introduced  for $1+2$ dimensional case find their interpretation: the ordinary duality takes us into the realm of wings around the dipyramid, while internal and external dualities carry us out of the light bipyramid.

We point out, however,  that possibilities for scators to be physically interpreted is strongly suppressed by the fact that the scator algebra does not possess rotational invariance. Fortunatelly,  the considered dynamics does not provoke the appearance of absolute-rest frame \cite{F1}, which leaves some hope for potential potential applications.

\end{document}